\documentclass[11pt,preprint]{aastex}
\pdfoutput=1
\usepackage{times}
\usepackage{graphicx}

\newcommand{\fbkg}{f_{\rm BKG}}
\newcommand\ltsima{$\; \buildrel <\over\sim \;$}
\newcommand\simlt{\lower.5ex\hbox{\ltsima}}
\newcommand\gtsima{$\; \buildrel >\over\sim \;$}
\newcommand\simgt{\lower.5ex\hbox{\gtsima}}
\newcommand\etal{et~al.}

\newcommand\msun {M_\odot}

\newcommand\ie{{i.e.}}

\def\etal{{\it et al.~}}
\def\secspt{$\buildrel{\prime\prime}\over .$}

\begin{document}

\shorttitle{HST CMD of Microlensing Source Stars}

\title{The MACHO Project HST Follow-Up: The Large Magellanic Cloud Microlensing
Source Stars}

\author{
      C.A.~Nelson\altaffilmark{1,2},
      A.J.~Drake\altaffilmark{3},
      K.H.~Cook\altaffilmark{1,4},
      D.P.~Bennett\altaffilmark{3,5},
       P.~Popowski\altaffilmark{6},
       N.~Dalal\altaffilmark{7},
       S.~Nikolaev\altaffilmark{1},
      C.~Alcock\altaffilmark{3,8},
    T.S.~Axelrod\altaffilmark{9},
      A.C.~Becker\altaffilmark{10},
    K.C.~Freeman\altaffilmark{11},
      M.~Geha\altaffilmark{12},
      K.~Griest\altaffilmark{13},
    S.C.~Keller\altaffilmark{1},
    M.J.~Lehner\altaffilmark{8,14},
    S.L.~Marshall\altaffilmark{15},
    D.~Minniti\altaffilmark{16,17},
    M.R.~Pratt\altaffilmark{18},
    P.J.~Quinn\altaffilmark{19},
    C.W.~Stubbs\altaffilmark{4,20},
      W.~Sutherland\altaffilmark{21},
    A.B.~Tomaney\altaffilmark{22},
      T.~Vandehei\altaffilmark{6},
      D.~Welch\altaffilmark{23}\\
      {\bf (The MACHO Collaboration)}
        }

\clearpage

\begin{abstract}
We present Hubble Space Telescope (HST) WFPC2 photometry of 13 microlensed 
source stars from the 5.7 year Large Magellanic Cloud (LMC) survey
conducted by the MACHO Project. The microlensing source
stars are identified by deriving accurate centroids in the ground-based MACHO
images using difference image analysis (DIA) and then transforming the DIA
coordinates to the HST frame. None of these sources is coincident with
a background galaxy, which rules out the possibility that the MACHO LMC
microlensing sample is contaminated with misidentified supernovae or AGN
in galaxies behind the LMC. This supports the conclusion that the
MACHO LMC microlensing sample has only a small amount of contamination
due to non-microlensing forms of variability. We compare the WFPC2 source star
magnitudes with the lensed flux predictions derived from microlensing fits
to the light curve data. In most cases the source star
brightness is accurately predicted. Finally, we
develop a statistic which constrains the
location of the Large Magellanic Cloud (LMC) microlensing
source stars with respect to the distributions of stars and dust in the LMC
and compare this to the predictions of various models of LMC
microlensing.  This test excludes at $\gtrsim
90$\% confidence level models where more than 80\% of the source stars lie
behind the LMC. Exotic models that attempt to explain the
excess LMC microlensing optical depth seen by MACHO with a population of
background sources are disfavored or excluded by this test. 
Models in which most of the lenses reside in a halo or 
spheroid distribution associated with either the Milky Way or the LMC are
consistent which these data, but LMC halo or spheroid models are favored
by the combined MACHO and EROS microlensing results.
\end{abstract}


\clearpage

\altaffiltext{1}{Lawrence Livermore National Laboratory, Livermore, CA 94550\\
    Email: {\tt kcook@igpp.ucllnl.org}}

\altaffiltext{2}{Department of Physics, University of California, Berkeley,
     CA 94720}

\altaffiltext{3}{Center for Advanced Computing Research, Caltech, 
Pasadena, CA 91125}

\altaffiltext{4}{Center for Particle Astrophysics, University of California,
	Berkeley, CA 94720}

\altaffiltext{5}{Department of Physics, University of Notre Dame, Notre Dame, IN 46556\\
    Email: {\tt bennett@nd.edu}}

\altaffiltext{6}{Max-Plank-Institut fur Astrophysik, 85741, Garching bei
Munchen, Germany}

\altaffiltext{7}{CITA, University of Toronto, 60 St. George St., Toronto, 
Ontario M5S 3H8, Canada 
}

\altaffiltext{8}{Harvard-Smithsonian Center for Astrophysics, Cambridge, MA 02138\\
	Email: {\tt calcock@cfa.harvard.edu}}

\altaffiltext{9}{Steward Observatory, The University of Arizona.
933 N. Cherry Ave, Tucson, AZ 85721}

\altaffiltext{10}{Department of Astronomy, University of Washington,
                 Box 351580, Seattle, WA 98195-1580
                 }

\altaffiltext{11}{Research School of Astronomy and Astrophysics,
        Mount Stromlo Observatory, Cotter Road, Weston, ACT 2611, Australia}

\altaffiltext{12}{Astronomy Department, Yale University, New Haven, CT~06520}

\altaffiltext{13}{Department of Physics, University of California,
    San Diego, CA 92039
    }

\altaffiltext{14}{Institute of Astronomy and Astrophysics, Academia Sinica.
P.O. Box 23-141, Taipei 106, Taiwan}

\altaffiltext{15}{SLAC National Accelerator Laboratory, Menlo Park, CA, 94025, USA}

\altaffiltext{16}{Department of Astronomy and Astrophysics, Pontificia 
Universidad Catolica de Chile, Casilla 306, Santiago 22, Chile
}

\altaffiltext{17}{Vatican Observatory, V00120 Vatican City State, Italy}

\altaffiltext{18}{Illumina Inc, 
              Hayward, CA 94545}

\altaffiltext{19}{School of Physics, University of Western Australia, Perth, Australia}

\altaffiltext{20}{Department of Physics, Harvard University,
                 Cambridge, MA 02138\\
    Email: {\tt stubbs@physics.harvard.edu}}

\altaffiltext{21}{Department of Physics, University of Oxford,
    Oxford OX1 3RH, U.K.
    }

\altaffiltext{22}{Pacific Biosciences Inc., 1505 Adams Drive,
                 Menlo Park, CA 94025}

\altaffiltext{23}{Department of Physics and Astronomy, McMaster University,
    Hamilton, Ontario, Canada, L8S 4M1
    }

\section{Introduction}

Gravitational microlensing was proposed by \citet{pac86} as a method to determine if
the dark halo of the Milky Way was comprised of objects such as Jupiters, brown dwarfs,
or white dwarfs in the planetary to stellar mass range. This proposed program has been
carried out by both the MACHO and EROS collaborations with remarkable success.
Objects ranging from $10^{-7}$ to $30\,\msun$ are excluding from dominating the
mass of the Milky Way's dark halo 
\citep{macho-eros-spike,macho_lmc5.7,macho_1_30_Msun,eros_lmc07}. This rules out
what were formerly the leading astrophysical dark matter candidates, and leaves
exotic particle dark matter as the clear leading candidate to comprise the 
dark matter halo of the Milky Way. However, MACHO collaboration has reported a
microlensing optical depth toward the central regions of the Large Magellanic Cloud
(LMC) that is substantially in excess of the expected microlensing
optical depth from known stellar populations \citep{macho_lmc5.7}. This implies that 
$\sim 20\,$\% of the Milky Way's dark halo could be in the form of massive compact
halo objects or MACHOs. The timescales of these microlensing events suggest that
these MACHOs could be low mass stars or white dwarfs. However, the EROS collaboration
has found a substantially lower microlensing optical depth \citep{eros_lmc07} with
a survey that focused on the outer regions of the LMC. When interpreted in terms of
lensing by a halo population, the MACHO and EROS results are in contradiction.

A number of ways to resolve this apparent contradiction have been considered. 
One possibility is an error by one of the groups. Within
the MACHO team, concern initially focused on the event detection methods and
the detection efficiency methods employed by EROS. Unlike MACHO
\citep{macho_eff}, EROS did not add simulated stars to images from their survey in
order to estimate their detection efficiency, and the EROS survey failed to find
a number the microlensing events detected by MACHO. However, simulated
image tests by EROS show that any systematic error in their efficiency
determination is not nearly large enough to explain the difference in the microlensing
optical depth measurements toward the LMC \citep{eros_lmc00,eros_lmc07}.
Furthermore, EROS has also shown that the LMC microlensing events that they
did not detect probably are properly accounted for by their detection
efficiency calculation.

There has also been some concern that the MACHO sample might be contaminated
by non-microlensing variability. This was suggested by \citet{belo_lmc_contamBS},
but their methods were criticized by \citet{griest_contam} and \citet{lmc_confirm}. However, 
it was later discovered by EROS \citep{eros_lmc07} that one of the MACHO
microlensing candidates (LMC-23) had a repeat variation. This could, in principle, be caused
by a binary source or lens, but neither bump on the light curve provides a very good
fit to a microlensing light curve, so it is most likely that this event, LMC-23, is a variable
star. The possibility of variable star contamination was considered in some detail by
\citet{lmc_confirm}, who found that improved difference imaging photometry of MACHO
data as well as CTIO follow-up data generally confirmed the microlensing interpretation
of the MACHO events. They also showed that the \citet{belo_lmc_contamBS} classified
LMC-23 and some background supernovae as microlensing events, while classifying
events with extremely good microlensing light curve fits as likely variables.
\citet{lmc_tau_cont} developed a method to correct the MACHO LMC optical depth
measurement for variable star contamination, and showed that this correction only
modestly reduced the MACHO LMC microlensing optical depth measurement.

Since experimental error by either MACHO or EROS seems unlikely, we are led to
consider astrophysical solutions to this apparent discrepancy. Since the LMC microlensing
rate is too small to be consistent with a dark halo comprised of only MACHOs, there
is no reason to assume that the lens objects trace the density distribution of the 
dark halo. However, the distinction between different potential microlensing
populations is complicated by the fact that the
crucial observable in microlensing, the event duration, typically admits degeneracy
in the three fundamental microlensing parameters: the mass, distance and
velocity of the lens.  This makes it difficult to distinguish between the two
principal geometric arrangements which may explain Large Magellanic Cloud
(LMC) microlensing: a) MW-lensing, in which the lens is part of the
Milky Way (MW) and b) self-lensing, in which the lens is part of the
LMC.  In self-lensing, the lens may belong to the disk+bar or spheroid
of the LMC, while the source star may come from either of these components or
some sort of background population to the LMC.

Most efforts to distinguish between MW-lensing and self-lensing focus on
modeling the LMC self-lensing contribution to the optical depth and comparing
this to the observed optical depth.  Since the observed optical depth is
substantially higher than the modelled self-lensing contribution for standard
models of the LMC, we have inferred 
that there must be a substantial MW-lensing signal \citep{macho_lmc5.7}, although
LMC self-lensing is certainly a possibility if we allow non-standard LMC models.

In this work, we take a different approach in which we discriminate between MW
lensing and self-lensing by locating the source stars and comparing their
location with predictions for various self-lensing and MW-lensing geometries.

We consider three types of LMC self-lensing: LMC disk+bar self-lensing, LMC
spheroid self-lensing and background lensing. Each type of self-lensing has a
distinct geometry (location of source star and lens), described in detail in 
\S~\ref{lmcmodels}.  The most important distinction is that the source star
populations are different for each self-lensing geometry.  In particular, in
each self-lensing geometry a different fraction, $\fbkg$, of the source stars
will lie behind the LMC disk.  The remaining fraction $(1.0-\fbkg)$ of the
source stars lie in the LMC disk+bar.  Therefore, if we determine the location
of the observed microlensing source stars we may compare the observed value of 
$\fbkg$ with the prediction for each model of LMC self-lensing.  In this way
we may eliminate some self-lensing geometries as possible explanations for the
LMC microlensing signal. We emphasize, that as defined in this paper, $\fbkg$
implies the fraction of source stars which lie behind the entire LMC disk, not
behind the mid-plane of the disk.

If all self-lensing geometries which satisfy other external constraints are
eliminated and the observed value of $\fbkg$ remains consistent with its
prediction in the MW-lensing geometry, we may conclude that the most likely
explanation for the microlensing signal is lensing by Massive Compact
Halo Objects (MACHOs) in the MW.

We determine the fraction of MACHO source stars drawn from a background
population lying behind the LMC disk, $\fbkg$,  using the suggestion of
\citet{zha99}, \citet{zha00a}, and \citet{zha00b}. These works point out that
source stars from a background population should be preferentially fainter and
redder than the average population of the LMC because they will suffer from
the extinction of the LMC disk, and are displaced along the line of sight.
\citet{zha00b} present a model for this background population with an
additional mean reddening relative to the average population of the LMC,
$\overline{E(B-V)} = 0.13$ mag, and a displacement from the LMC of $\sim 7.5$
kpc.  The displacement results in an increase of the distance modulus by
$\overline{\Delta \mu} \sim 0.3$ mag. 

We note that, in reality, even disk+bar self-lensing would have a
source star distribution which is redder than for MW-lensing.  In
MW-lensing, the source stars will be distributed evenly throughout the LMC
disk.  In the disk+bar self-lensing disk there will be more source stars on
the far side of the LMC disk (e.g., behind the mid-plane) than on the near
side of the LMC disk.  The fraction of far-side source stars depends on your
model for the LMC disk and the reddening effect depends sensitively on your
assumptions about the distribution of dust in the LMC (e.g., the dust scale
height).  This level of extra reddening is a much more subtle effect, and much
more model dependent.  We do not address this situation here, instead we only
consider the reddening effects of source stars lying behind the entire LMC
disk.

We test for possible extra reddening in our source stars relative to the
average population of the LMC by comparing a color-magnitude diagram (CMD) of
the source stars with the CMD of all nearby LMC stars.  The CMDs are
created using Wide Field Planetary Camera 2 (WFPC2) Hubble Space Telescope
(HST) photometry of the 13 microlensing source stars selected using cut A of
\citet{macho_lmc5.7} and their surrounding fields.  The source stars are identified by deriving accurate
centroids in the ground-based MACHO images using difference image analysis
(DIA) and then transforming the DIA coordinates to the WFPC2 frame.  We compare
the source star color magnitude diagram (CMD) with a 2-D
Kolmogorov-Smirnov (KS) test to model source star CMDs in which varying
fractions of the source stars are located in and behind the LMC. 

In \S~\ref{lmcmodels} we describe the four models of LMC microlensing. In
\S~\ref{hstobs} we construct a CMD of the average LMC population by combining
the CMDs of thirteen WFPC2 fields centered on
past MACHO microlensing events in the outer LMC bar.  In \S~\ref{identify} we
describe the identification of the microlensing source stars in these fields
by difference image analysis, and find that there are no background galaxies
in the vicinity of the source stars. In \S~\ref{blend_fits} we show that the source
star brightness implied by the microlensing fits matches the source star
observed in the HST images for most events and discuss the reasons why
the fit magnitudes occasionally fail to match the observed magnitudes.
In \S~\ref{modelpops} we describe the
construction of model source star CMDs with varying fractions of background
source stars, $\fbkg$.  In \S~\ref{stats} we determine the likelihoods that
the observed microlensing source stars were drawn from each of the model
source star CMDs by using Kolmogorov-Smirnov (KS) tests to compare the
observed and model distributions. We discuss the results of the
KS test in \S~\ref{ssdiscuss},  in the context of four models of
microlensing: a) MW halo-lensing, b) LMC disk+bar
self-lensing, c) LMC spheroid self-lensing, and d) background lensing.
We also consider other constraints on these microlensing models, such as
the microlensing results from the EROS project and events which have lens
distance determinations. Finally, in \S~\ref{sssum}, we summarize our results.

\section{Models of LMC microlensing}
\label{lmcmodels}

The most straightforward interpretation of the LMC microlensing signal is that
the source stars in the disk+bar of the LMC are being lensed by MACHOs in the
Milky Way halo.  In this work we refer to this geometry as MW-lensing.  Since all
the source stars are located in the LMC disk+bar, the expected value of
$\fbkg$ is $\sim 0.0$.  The other models we consider are all variants on LMC
self-lensing, meaning that the lenses themselves also belong to some component
of the LMC.

Early works by \citet{sah94} and \citet{wu94} considered lensing by the LMC
in slightly different contexts. \citet{wu94} considered lensing by the dark halo
of the LMC as well as known stellar populations and showed that the LMC dark
halo should have a microlensing optical depth that is slightly larger than the
optical depth measured by MACHO. He also argued that lensing by LMC stars
could be important. \citet{sah94} argued that the entire LMC microlensing signal
could be explained by ordinary stellar LMC populations, so that no dark matter lensing
was needed. In this work, we refer to this geometry as LMC disk+bar
self-lensing.  Both these early works suggested that LMC disk+bar self-lensing
could account for a substantial fraction of the observed optical depth.  This
claim has since been disputed by several other groups
\citep{gou95,alc97,macho_lmc5.7,mancini_lmcdisk} 
who show that when considering only disk stars
the LMC self-lensing optical depth is far too low to account for the observed
optical depth.  \citet{gyu00} show that allowing for
contributions to the lens and source populations from the LMC bar does not
substantially increase the LMC self-lensing optical depth. \citet{alv00a}
find a low LMC self-lensing optical depth for a flared LMC disk.
\citet{gyu00} also show that much of the disagreement in models of
the LMC disk+bar self-lensing optical depth results from disagreement about
the fundamental parameters of the LMC, such as the total disk mass and
inclination angle.  Within their region of allowed parameters \citet{gyu00}
also make a strong case that LMC disk+bar self-lensing makes a
small contribution to the observed optical depth, at most $\sim$25\%
(using the MACHO result, $\tau_{\rm LMC} = (1.0\pm 0.3) \times 10^{-7}$, corrected for 
variable star contamination \citep{lmc_tau_cont}). \citet{mancini_lmcdisk} show
that the spacial distribution of microlensing events indicates that a large
fraction of them are not caused by the LMC disk+bar.

In the LMC disk+bar self-lensing model, the source stars will be
preferentially drawn from the far side of the LMC disk (e.g., beyond the
midplane of the disk), but are still part of the LMC disk and are not drawn
from a
distinct entity behind the entire LMC disk.  Therefore, this model also gives
$\fbkg \sim 0.0$.

Although the upper limit on the optical depth expected from LMC disk+bar
self-lensing is strongly constrained to be substantially below the observed
optical depth, the self-lensing optical depth may be raised much higher if one
allows for lenses in an LMC stellar halo population as first shown by
\citet{wu94}.  The principal problem
surrounding an LMC stellar halo contribution is that tracers of old
populations in the LMC seem to indicate that this population has too little mass
to contribute more than $\sim 10$\% of the optical depth measured by MACHO.

However, another possibility has arisen from the results of
\citet{wei00} who claims that LMC microlensing may be caused by a
\textit{non-virialized} stellar halo or ``shroud''. In this geometry the
lens may lie in the LMC disk+bar or the foreground part of the shroud, while
the source stars lie in the LMC disk+bar or the background part of the shroud.
The term shroud was introduced
by \citet{eva00} and is meant to imply an LMC population which is
like a halo in that it is spatially not part of the LMC disk, but unlike a
halo in that it is non-virialized and thus may have a relatively low velocity
dispersion.  Such a population is suggested by the simulations of
\citet{wei00}, who finds that the LMC's dynamical interaction with the MW may torque
the LMC disk in such a way that the LMC disk is thickened and a spheroid
component is populated without isotropizing the stellar orbits and thereby
leaving disklike kinematics intact. 

However, observational work on RR Lyrae by \citet{kin91}, which does not
rely on the specific kinematics of the spheroid, limits the total mass of any
type of halo (virialized or non-virialized) to perhaps 5\% of the mass of the
LMC, too small to contribute more than $\sim$5\% of the observed optical
depth.  A more recent revisiting of this argument by \citet{alv00b} 
finds room to increase this optical depth contribution to at most 20\%, still
only a small fraction of the total. The \cite{wei00} models have been
challenged on various grounds by \cite{kui01}.

In order for an LMC shroud to account for the total optical depth it must have
a mass comparable to that of the LMC disk+bar \citep{gyu00}.
Even if we accept the existence of such a massive shroud, the microlensing
implications are somewhat in dispute.  \citet{wei00} finds an LMC
self-lensing optical depth comparable to the observed optical depth.  However,
this estimate is reduced by a factor of 3 by \citet{gyu00} 
who repeat the \citet{wei00} microlensing analysis using lower values for
the disk total mass and inclination angle, and a proper weighting over all
observed MACHO fields.

In this work we refer to both LMC halo and shroud self-lensing as LMC spheroid
self-lensing.  In spheroid self-lensing there are four event geometries: a)
background spheroid source and disk lens, b) disk source and foreground
spheroid lens, c) disk source and disk lens and, d) background spheroid source
and foreground spheroid lens.  We might naively expect the former two types
dominate the number of expected events, and so if we were to ignore the
contribution from the latter two types we would conclude that spheroid lensing
would imply $\fbkg \sim 0.5$.  However, in order to produce the entire
observed optical depth, the spheroid must be so massive that it is no longer
self-consistent to ignore the final term.  Calculations performed in the
formalism of \citet{gyu00} suggest instead that events with a
background spheroid source and a foreground spheroid lens become an important
contributor and increase the expected fraction of background source stars to
$\fbkg \sim 0.65$. However, this estimate assumes that the spheroid has the same
stellar population as the disk+bar. If the spheroid population is much older, it could
lack the main sequence A and F stars that serve as sources for 75\% of the 
LMC microlensing events. In this case, a much smaller value of $\fbkg$ would be
appropriate.

Yet another self-lensing geometry was introduced by \citet{zha99} who suggests
that the observed events are due to ``background'' self-lensing in which the
source stars are located in some background population, displaced at some
distance behind the LMC. A veritable plethora of lenses for this population is
then supplied by the disk+bar of the LMC. A background population has the
advantage of being nearly impossible to confirm or reject observationally, as
there are nearly no limits on its size or content (provided of course, it is 
small enough to ``hide'' behind the LMC). Since all the source stars belong to
the background population, $\fbkg \sim 1.0$.

\begin{figure}
\epsscale{0.8}
\plotone{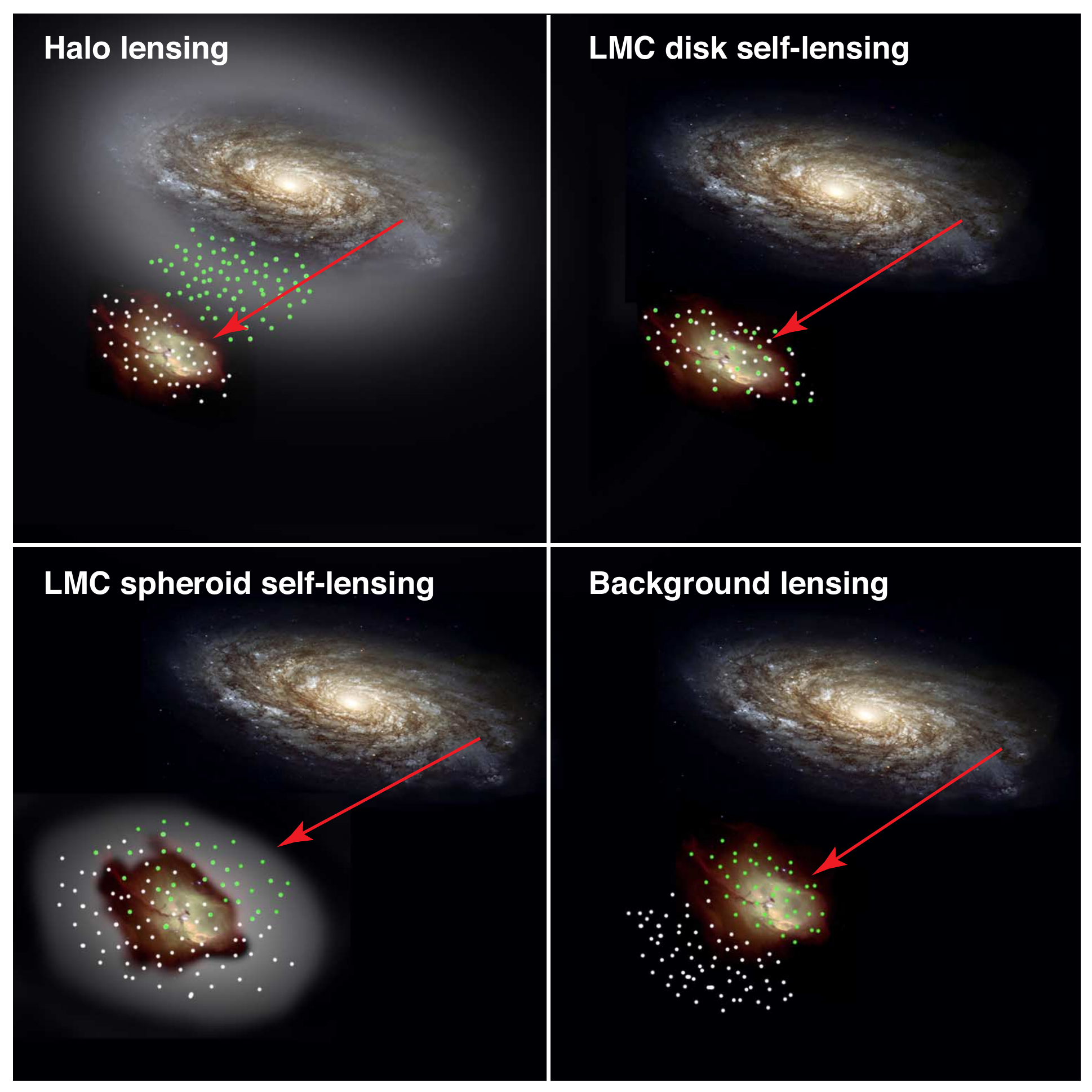}
\caption{Possible LMC microlensing geometries.
The Milky Way is depicted as the large spiral galaxy in the upper
right hand corner, the LMC as the irregular galaxy in the lower left hand
corner. In each figure the arrow indicates a line of sight from the location
of the Earth in the Milky Way towards a random position in the LMC. The white dots
indicate the position of the source stars, the green dots the position of the
lenses. Figure not to scale.}
\label{modelpictures}
\end{figure}

Each of the above LMC microlensing geometries is depicted pictorially in
Figure~\ref{modelpictures}.

A final possibility is foreground self-lensing.  This is not self-lensing
in the classical sense as in this case the lenses are not drawn from the LMC
itself, but rather from some kinematically distinct foreground population,
such as an intervening dwarf galaxy.  \citet{zar97} claim a detection
of a population of stars from such an entity.  However, \citet{bea98} 
claim that this population is a morphological feature of the LMC
red clump, while others show that such a population consistent with other
observational constraints could not produce a substantial microlensing signal
\citep{gou98,ben98}.  \citet{zar_etal_99} argue that may be possible to evade these
constraints by tuning the IMF of this putative foreground population. But the
{\it a priori} probability for such a large mass of stars to exist in the foreground
of the LMC is already quite low, so this explanation seems rather unlikely.
We do not consider this model any further. 

\section{WFPC2 Observations of Microlensing Source Stars}
\label{hstobs}

Observations were made under proposals GO-05901, GO-7306 and GO-8676
(Principal Investigator: Kem Cook) with the WFPC2 on the HST between May 1997
and July 2000 through the F555W ($V$) and F814W ($I$) filters.  The Planetary
Camera (PC) was centered on the location of past MACHO microlensing events.
The microlensing events, positions, and exposure times are listed in Table 1.

Multiple exposures of a field were combined using a sigma-clipping algorithm
to remove deviant pixels, usually cosmic rays.  The PC has a pixel size of
0\secspt046 which easily resolves the great majority of stars in our frames.
Most stars are also resolved in the Wide Field (WF) fields which have a pixel
size of 0\secspt1.  Instrumental magnitudes were calculated from aperture
photometry using DAOPHOT II \citep{ste87,ste91} with a radius of 
0\secspt25 and
centroids derived from point-spread function (PSF) fitting photometry.
Aperture corrections to 0\secspt5 were performed individually for each frame. 
We choose this aperture radius as this is how the \citet{hol95} 
calibrations are defined. We
correct for the WFPC2 charge transfer effect using the equations from
Instrument Science Report WFPC2 97-08.  We also make the minimal corrections
for contaminants which adhere to the cold CCD window according to Table 28.2
of the HST Data Handbook.  We transform our instrumental magnitudes to Landolt $V$
and $I$ using the calibrations from \citet{hol95}.

\begin{figure}
\epsscale{1.0}
\plotone{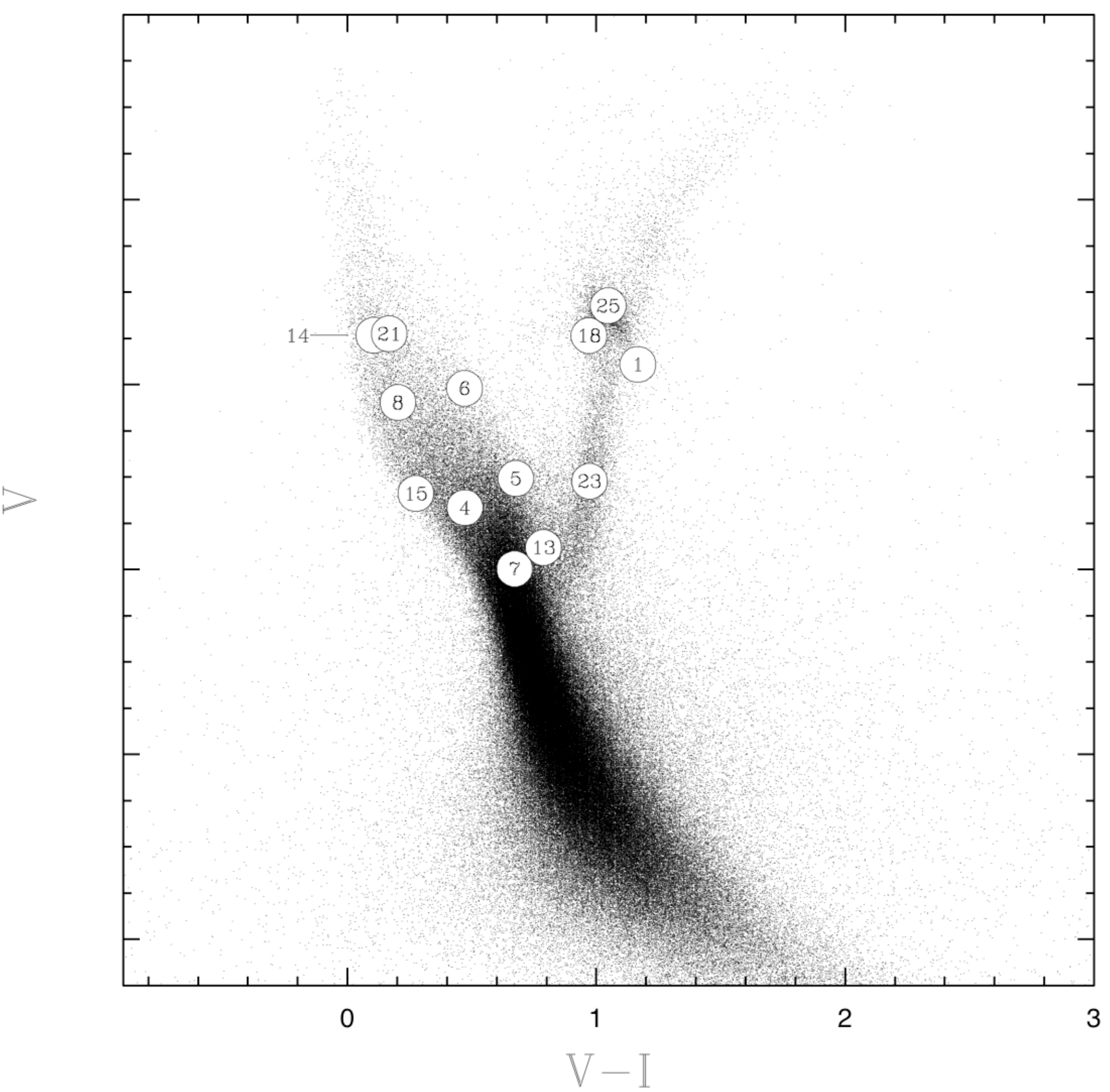}
\caption{Color magnitude diagram of LMC microlensed
  source stars. The source stars are overplotted on a 
  composite HST CMD of the fields surrounding each of these events.
  The numbers indicate the corresponding MACHO microlensing event.
  }
\label{cmd}
\end{figure}

We create a composite LMC CMD by combining the PC and WF photometry for all of
our fields except the field of LMC-1.  In the case of LMC-1 the $V$ and $I$
observations were taken at different roll angles and there is little area of
overlap except in the PC frame.  We therefore include the PC field from LMC-1
but not the WF fields.  The composite HST CMD is shown in Figure~\ref{cmd}.

\section{Source Star Identification}
\label{identify}

A ground-based MACHO image has a pixel size of 0\secspt6 and a seeing of at
least 1\secspt4.  In a typically crowded region of the outer LMC bar, a MACHO
seeing disk will contain $\sim$11 stars of $V\lesssim24$.  This means that
faint ``stars'' in ground-based MACHO photometry are usually not single stars
at all, but rather blended composite objects made up of several fainter stars.
Henceforth, we distinguish between these two words carefully, using
\textit{object} to denote a collection of stars blended into one seeing disk,
and \textit{star} to denote a single star, resolved in an HST image or through
difference image analysis.  The characteristics of the MACHO object that was
lensed tell us little about the actual lensed star.  However, with the
microlensing object centroid from the MACHO images we can hope to identify the
microlensing source star in the corresponding HST frame.

\begin{figure}
\plotone{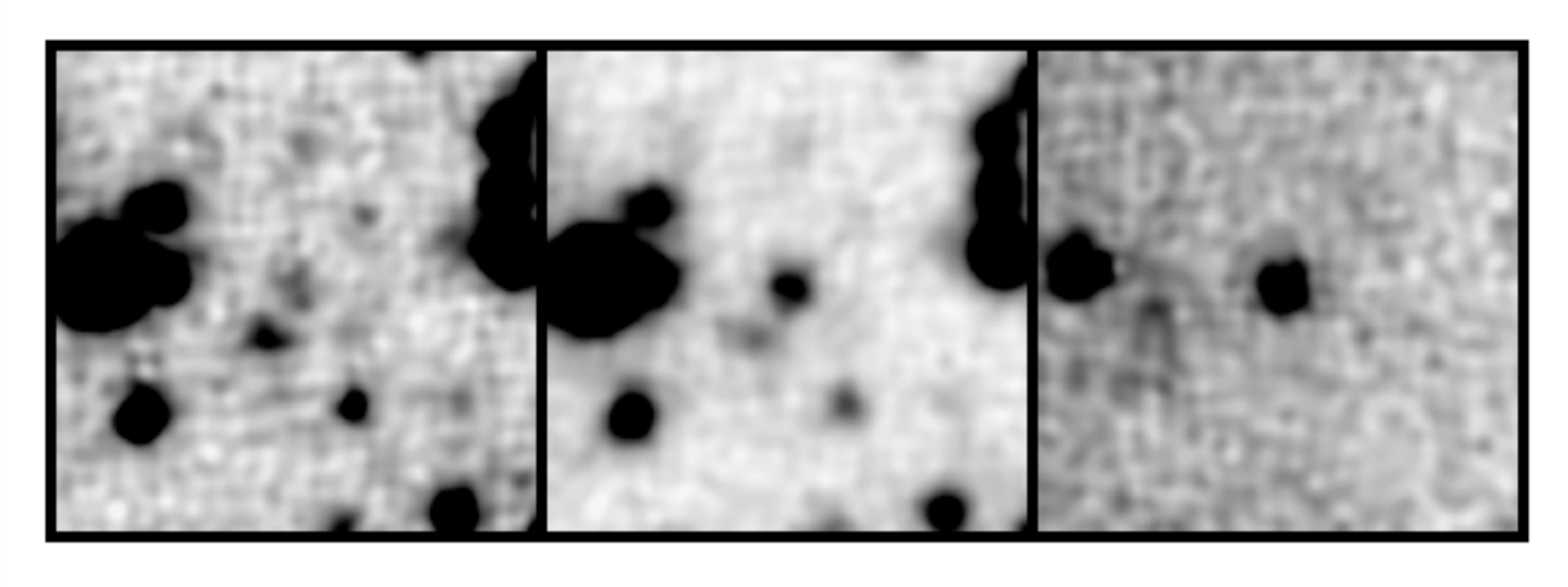}
\caption{Difference Image Analysis.
  The left panel shows a $0.6^{\prime}~\times~0.6^{\prime}$ section of the
  baseline image of MACHO event LMC-4. The middle panel
  shows the same region taken at the peak of the microlensing event.  The 
  right panel shows the differenced image. The flux in the middle of the
  differenced image is the extra flux from the microlensing event. The flux at the left hand
  side of the differenced image is an expected contamination due to an asymptotic giant branch variable
  star at that location. }
\label{macho}
\end{figure}

\begin{figure}
\epsscale{0.5}
\plotone{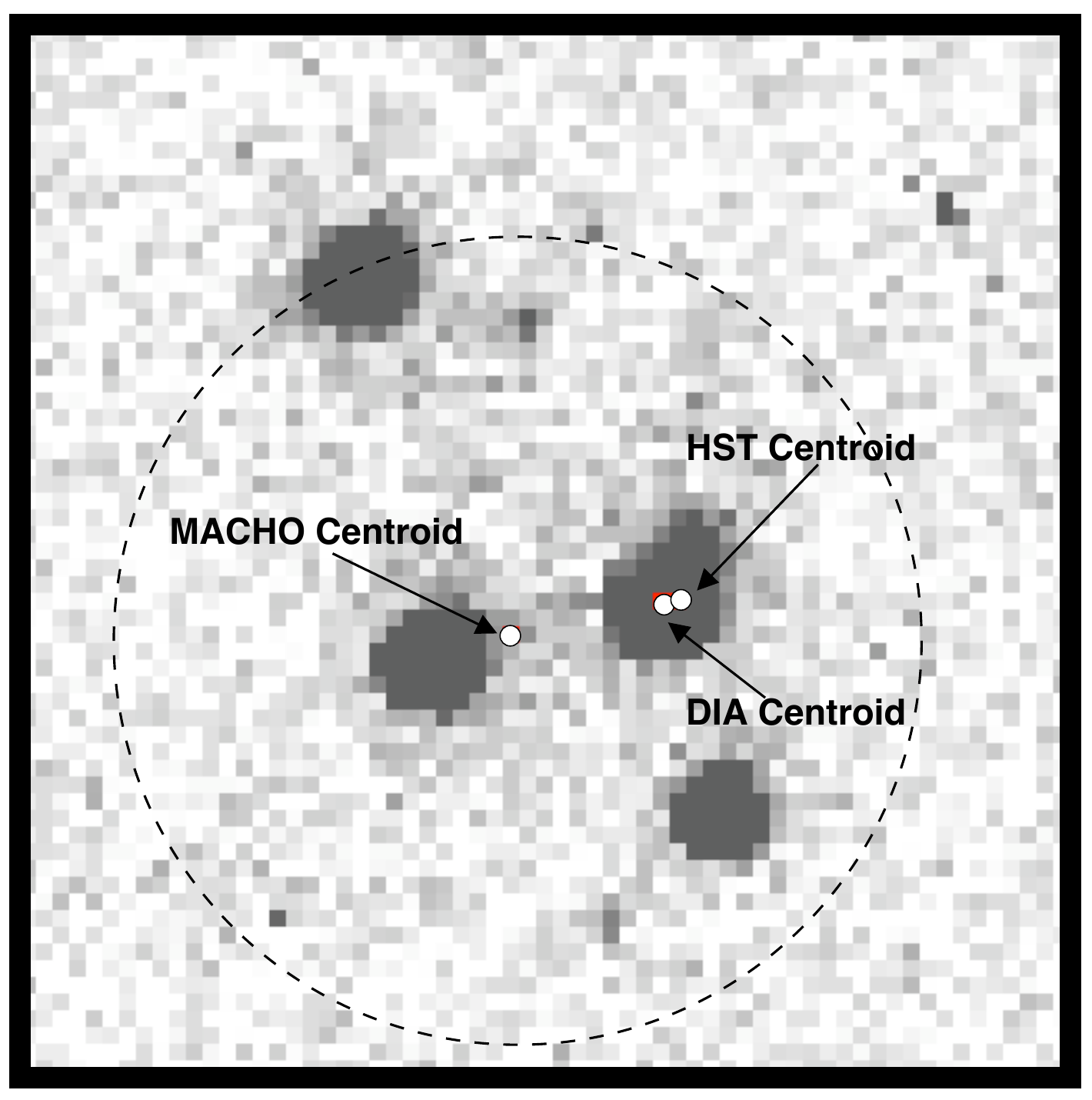}
\caption{DIA results for event LMC-4.
A $3^{\prime\prime}~\times~3^{\prime\prime}$ HST image of LMC-4.
The circle contains the
several HST stars which are all contained within one MACHO seeing disk of the
lensed object.  The arrows indicate the MACHO baseline centroid, the DIA
centroid and the centroid of the HST star nearest to the DIA centroid. }
\label{lmc4}
\end{figure}

A direct coordinate transformation from the MACHO frame to the HST frame often
places the baseline MACHO object centroid in the middle of a group of faint
HST stars with no single star clearly identified.  To resolve this ambiguity we
have used DIA.  This technique is described in detail by \citet{tom96} and
\citet{alc99}, but we
review the main points here.  DIA is an image subtraction technique designed
to provide accurate photometry and centroids of variable stars in crowded
fields.  The basic idea is to subtract a flux and seeing matched high S/N
reference image from each from each program image, leaving a differenced image containing
only the variable components. Applied to microlensing, we subtract baseline
images from images taken at the peak of the microlensing light curve, leaving
a differenced image containing only the flux from the microlensing source
star and not the rest of the object. We typically observe a centroid shift between the
MACHO object's location and the residual lensed flux in the difference image.
If the centroid from the differenced image is transformed to the
HST frame we find that it usually clearly identifies the HST microlensed
source star.  This process is illustrated in Figures~\ref{macho} and
\ref{lmc4}.

\begin{figure}
\epsscale{0.8}
\plotone{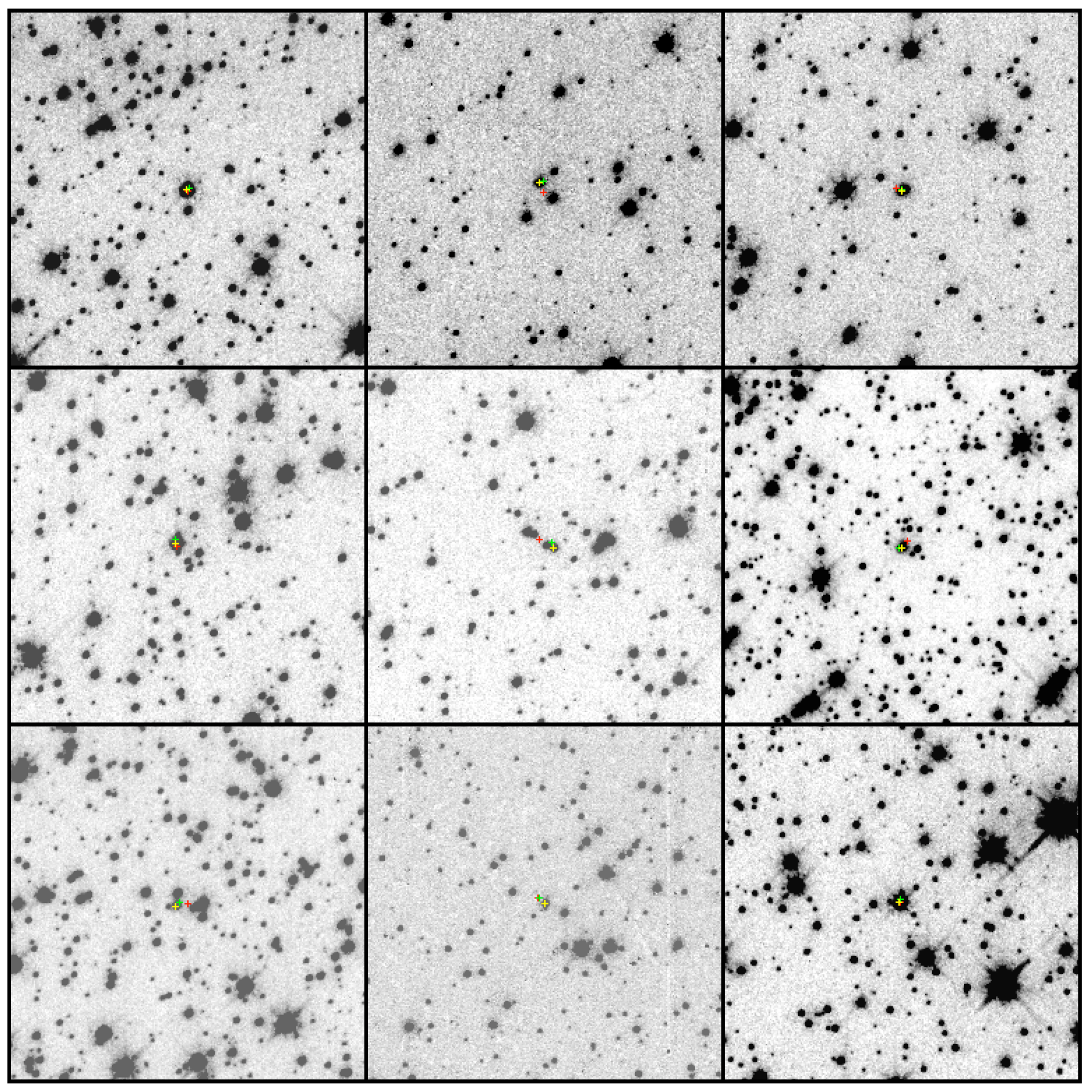}
\caption{HST finding charts for microlensing source stars I. From top to
bottom, left to right, we show LMC-1, LMC-4, LMC-5, LMC-6, LMC-7, LMC-8,
LMC-9, LMC-13, and LMC-14. The MACHO object centroid is shown as a red cross,
the DIA centroid is shown as a green cross and the centroid of the HST star
closest to the DIA centroid is shown as a yellow cross.} 
\label{dia1}
\end{figure}

\begin{figure}
\epsscale{0.8}
\plotone{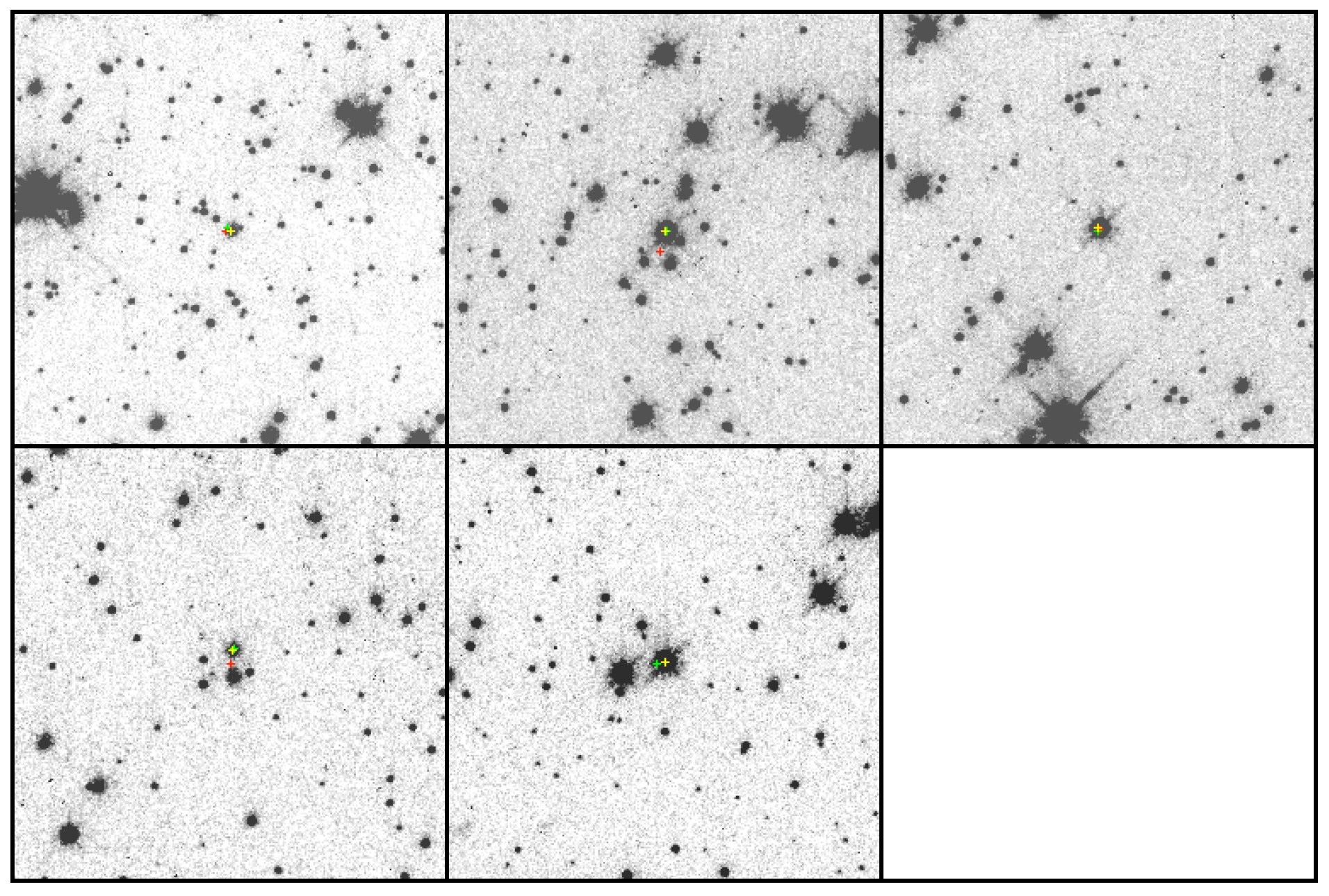}
\caption{HST finding charts for microlensing source stars II. 
As in Figure~\ref{dia1} for events LMC-15, LMC-18, LMC-21, LMC-23,
and LMC-25.}
\label{dia2}
\end{figure}

In Figures~\ref{dia1} and~\ref{dia2} we present WFPC2 finding charts
identifying the MACHO object centroid, the centroid in the DIA image and the
centroid of the star in the WFPC2 image closest to the DIA centroid.

This technique allows us to unambiguously identify the 13 microlensed source
stars which pass the cut A microlensing tests in \citet{macho_lmc5.7}.  In
Table 2 we present the $V$, $R$, and $I$ magnitudes of our source stars
from the HST data.  The errors presented here include the formal photon counting
errors returned by DAOPHOT II as well as an
additional $0.03$ mag uncertainty due to aperture corrections.  We note
that the cut A microlensing events do not include LMC-9 (which
appeared in a similar analysis by \citet{macho_hst1}).  Although LMC-9 is
certainly a valid microlensing event, it is a binary microlensing event 
\citep{macho-lmc9} and
therefore its non-standard light curve does not pass the stringent cut A
requirements.  Although we list its photometric information in Table 2, it is
not used in our statistical analysis.

Our identification of LMC-5 revealed it to be the rather rare case of a
somewhat blended HST star \citep{lmc5-detect}.  Although there are two stars evident, at an
aperture of 0\secspt25 the flux of one star was contaminated by that of its
neighbor. Therefore, in this case, we perform PSF fitting photometry using
PSFs kindly provided by Peter Stetson.  The errors presented in Table 2 for
LMC-5 are those returned by the profile fitting routine ALLSTAR. The DIA
centroid falls 2 pixels closer to the centroid of star one than star two,
clearly preferring star one as the source star. Furthermore, as predicted by
\citet{alc97} and \citet{gou97}, star two is a rather red object which is very
faint in the V band.  Fits to the MACHO light curve presented by
\citet{macho_lmc5.7}
suggest lensed flux fractions in the V and R bands of 1.00 and 0.46
respectively, confirming the DIA choice of the much bluer star as the lensed
source star. In addition to these HST/WFPC2 observations \citet{lmc5-detect},
a second epoch of HST observations using the Advanced Camera for
Surveys (ACS) were obtained \citep{drake_lmc5}, and these data combined
with a microlensing parallax analysis \citep{gould-jerkpar} resulted in the
first stellar mass measurement of a single star besides the Sun
\citep{lmc5-mass}. The LMC-5 lens star has a mass of $0.097\pm 0.016\msun$.

An important source of background for microlensing events in these crowded fields
is due to distant galaxies behind the LMC. This was first noticed in \citet{alc97},
where event LMC-11 was seen to be
very close to an obvious background spiral galaxy.
This and the asymmetric light curve indicated that the variability was almost
certainly due to a background supernova and not microlensing.  Background
supernovae were considered in some detail in \citet{macho_lmc5.7}, where a
detailed supernova rejection procedure was developed. Events were rejected
if
\begin{enumerate}
\item a supernova type Ia model provided a better fit than microlensing, or
\item a background galaxy was visible in the proximity of the event.
\end{enumerate}
This procedure is expected to work because the type II supernovae with light curves
that don't resemble type Ia light curves are generally associated with galaxies
that are closer and have more young stars, so that they should be visible in
the MACHO images. This procedure seems to remove all the supernovae 
contamination from the MACHO sample, but there are a few events, such as
LMC-10 and 22, for which host galaxy is so compact that it looks like a star
in the MACHO images. Thus, an important check of our background rejection
procedure is that all of the events should not be associated with background 
galaxies. A careful inspection of the HST images (see Figs.~\ref{dia1} and 
\ref{dia2}) reveals no extended sources in the vicinity of the lens events, so
the efficiency of  our supernovae rejection method is confirmed.

\section{Comparison with MACHO Blend Fits}
\label{blend_fits}

As discussed above, in a ground-based MACHO image each microlensed ``star''
is often a blended composite image of several fainter stars.  Only
one of the stars which make up the composite object is microlensed. In this
section, we have described one method for identifying the microlensed star in
the WFPC2 image.

However, there exists another method
for estimating the percentage of the total object flux which was
actually microlensed.  The ground-based object flux may separated into lensed and unlensed
components by performing a microlensing blend fit to the ground-based
lightcurve.  This procedure is described in detail by \citet{alc97} and
\citet{macho_lmc5.7}.

\begin{figure}
\epsscale{0.8}
\plotone{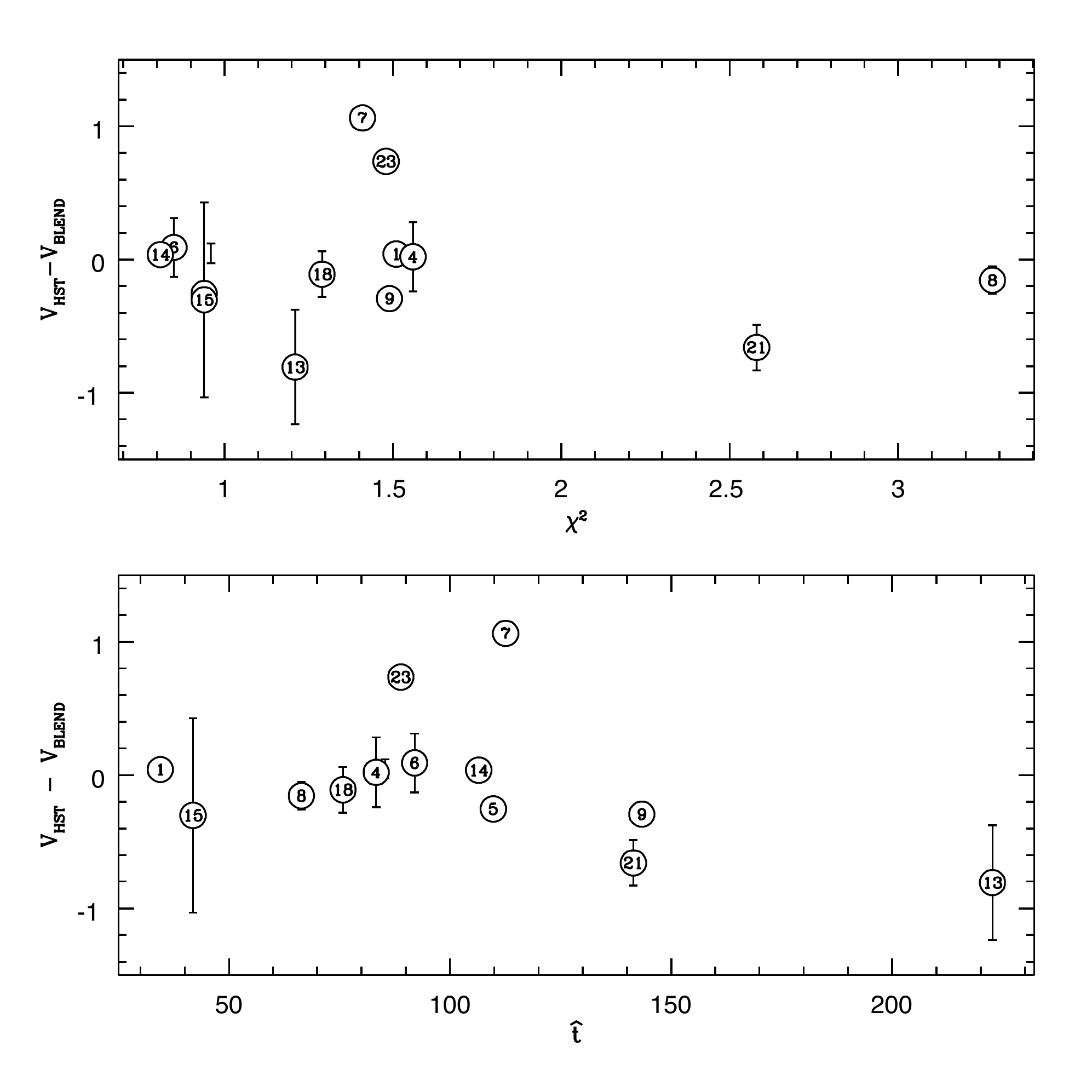}
\caption{A comparison between the HST source star magnitude, $V_{\rm HST}$
and the estimated source star magnitude derived from a microlensing blend fit
to the ground-based lightcurve, $V_{\rm blend}$. We plot the difference
versus the event-duration, $\hat{t}$ and the reduced chi-squared of the fit,
$\chi^{2}$. We show error bars only when they are larger than the dots.} 
\label{blend}
\end{figure}

In Table 2, we calibrate the lensed flux for each event as presented in Table 5
of \citet{macho_lmc5.7}.  The calibrated lensed flux is an estimate of the 
source star (not object) magnitude, derived entirely from the ground-based photometry.
Thus, we would expect $V_{\rm HST}\sim V_{\rm blend}$ and 
$R_{\rm HST}\sim R_{\rm blend}$.  We show a plot of the difference
$V_{\rm HST}-V_{\rm blend}$ in Figure~\ref{blend}.  We plot the
difference versus the event duration, $\hat{t}$ and the reduced chi-squared
of the microlensing blend fit (see Table 5 of \cite{macho_lmc5.7}).

The blend fits work reasonably well for all events except LMC-7, LMC-13, LMC-21
and LMC-23.  The case of LMC-13 is considered by \citet{lmc_confirm}, who show 
that the blend fit is consistent with the source identified in the HST frames when follow-up
CTIO photometry and improved DIA photometry of the MACHO survey data is used
in the fits. For LMC-21, the blend fits indicate a source that is $\sim 0.7$ mag fainter
than the corresponding source in the HST frames, but the $R-V$ colors of the blend fit
sources and the HST image are equal to within their errors. Thus, this event might be
explained if the source has a binary companion of almost equal brightness at a separation
larger than a few AU.

For LMC-7, the blend fit source also has a color that is similar to the color seen in the
HST images, but in this case the blend fit source is brighter. So, this cannot be explained
by a source companion. The light curve is symmetric and achromatic, with a peak magnification
of 6.9, but it also has significant deviations from the standard, single lens microlensing light
curve. This could be caused by a non-standard microlensing effect, such as a weak
binary lens, microlensing parallax \citep{gould-par1,macho-par1} or xallarap
\citep{griest_hu,multi-par}, which is the result of orbital motion of the source (\ie parallax reversed). 
The HST star for this event does appear to be slightly redder than the blend fit source. 
If this does not reflect photometry problems (as was the case with LMC-13), it might 
indicate a very red foreground Milky Way disk lens as was the case for events LMC-5
\citep{lmc5-detect} and LMC-20 \citep{kal_lmc20}. (LMC-20 is a lower S/N microlensing
event that did not pass all the sample A cuts.) If this excess
red flux is really from the source, then there is a very good chance that the light curve
will show a parallax deviation.

LMC-23 is the one source that shows a large color offset between the blend fits and the 
HST photometry, with the HST blend fits indicating a source about 0.3 mag bluer than the
HST frame.  This event also has significant light curve deviations that can probably
not be explained by a reasonable microlensing model \citep{lmc_confirm}. In fact, data 
from the EROS \citep{eros_lmc07} and OGLE (Udalski, 2004, private communication)
groups show an additional brightening that occurred after the MACHO survey stopped 
taking data. It is possible, but unlikely, 
for binary source or binary lens model to produce two light curve
bumps separated by a year or more \citep{macho-binaries}, but it is virtually impossible
to also ascribe the poor microlensing fit to non-standard microlensing. The only
reasonable conclusion is that this event is not microlensing, but is actually
 due to a variable star.
However,
this is the only MACHO LMC event that has been confirmed to be something besides a
microlensing event, and a careful analysis of all the MACHO LMC events indicates
that contamination of the MACHO sample by non-microlensing variability leads to
only a small correction to the microlensing optical depth \citep{lmc_tau_cont}.

\section{Creation of Model Source Star Populations}
\label{modelpops}

We consider source stars drawn from two distinct populations: the disk+bar of
the LMC and a background population located behind the LMC disk.  Source stars
from the disk+bar of the  LMC are well represented by our composite HST CMD
shown in Figure~\ref{cmd}.  Source stars drawn from the background population
will be on average redder than the disk+bar by an amount equal to the mean
extinction through the LMC and may also be displaced behind the LMC by
some distance.

The mean internal extinction of the LMC has been determined using UBV and UBVI
photometry of young, hot (O--A  type) stars in  \citet{oes96} and \citet{har97}. \citet{oes96}
find a mean internal reddening due to dust in the LMC disk of
$\overline{E(B-V)}=0.16$ and \citet{har97} find $\overline{E(B-V)}=0.13$.  

A more recent study \citep{zar99} suggests that the reddening of the LMC may be
population dependent. \cite{zar99} find $\overline{E(B-V)}=0.17$ for hot,
young stars ($T>22,000$ K), but a much lower internal reddening for older,
cooler stars ($5500~K<T<6500~K$) of $\overline{E(B-V)}=0.03$.  The lower
reddening for older stars may be due to the difference in scale heights of an
old population and a young population relative to the scale height of the dust
(i.e., the old population may have a larger scale height) combined with the
presence of low-extinction ``holes'' in the dust layer.  In this interpretation,
the large scale height of old stars implies that about half the cool stars
live in front of the LMC dust plane and are only extincted by Galactic foreground dust.
In contrast, the small scale height of the young stars implies that the
majority live within the dust plane and suffer from both foreground 
and internal extinction.
Of the stars lying behind the LMC midplane, some fraction are found in
``holes'' and are again only extincted by the Galactic foreground dust.
According to \citet{zar99} such a model works well in predicting both the
hot star and cold star extinctions in the LMC.

If the above interpretation of population dependent reddening is correct,
then the  mean reddening for the young hot stars is an appropriate measure
of the total internal reddening of the LMC disk.  However, the concept
of a patchy LMC extinction is problematic for our Kolmogorov-Smirnov test in
which we assume that the reddening of all of our fields is approximately the
same.  Empirically, we find that the reddening of our fields is approximately
the same.  If we shift our CMDs up and down the reddening axis until they
seem to line up perfectly (by eye), we find that the maximum shift is 
within $\Delta E(B-V) \sim 0.03$ of the mean.  

The high reddenings for young, hot stars may also be due to environmental
effects, e.g., the young stars may still live very close to the dust in which
they were formed.  In this case, the young star reddening may significantly
overestimate the true mean internal reddening of the LMC disk.

In this work, we adopt the same value as used in \cite{zha00b}, $\overline{E(B-V)}=0.13$.
This is the lowest mean value determined using young stars, but is
significantly higher than the \citet{zar99} value for old stars.
We note, that if the true value is significantly lower than
$\overline{E(B-V)} = 0.13$, the statistical significance of our 
Kolmogorov-Smirnov test would decrease substantially.

The distance
to the background population of $\Delta \mu \sim 0.3$ from \citet{zha00b} is
very loosely derived by the requirement that the background population be at
least transiently gravitationally bound to the LMC.  We consider three
different displacement distances from $\Delta \mu = 0.0$, $\Delta \mu = 0.3$
($\sim7.5$ kpc) and $\Delta \mu = 0.45$ ($\sim11.5$ kpc). We have no
constraints on the location, size or content of a possible background population, except that
it must be small enough and similar enough to the LMC stellar population to
have avoided direct detection.

For each model source star population we begin with the composite HST CMD and
then shift some fraction, $\fbkg$, of the stars into the background.  That is,
we redden and displace a fraction $\fbkg$ of the CMD  by amounts
\begin{equation}
\Delta V = A_V + \Delta \mu~,
\end{equation}
\begin{equation}
\Delta(V-I) = E(V-I) =
1.376\cdot E(B-V) = 0.18~{\rm mag}.
\end{equation}
The coefficient in the conversion from $E(B-V)$ to $E(V-I)$ is drawn from
Table 6 of \citet{sch98} and $A_V = 3.315\cdot E(B-V) =
0.43$ mag.  We use the Landolt coefficients from \citet{sch98} 
since the \citet{hol95} coefficients calibrate the WFPC2 magnitudes to
this system.  

\begin{figure}
\epsscale{0.5}
\plotone{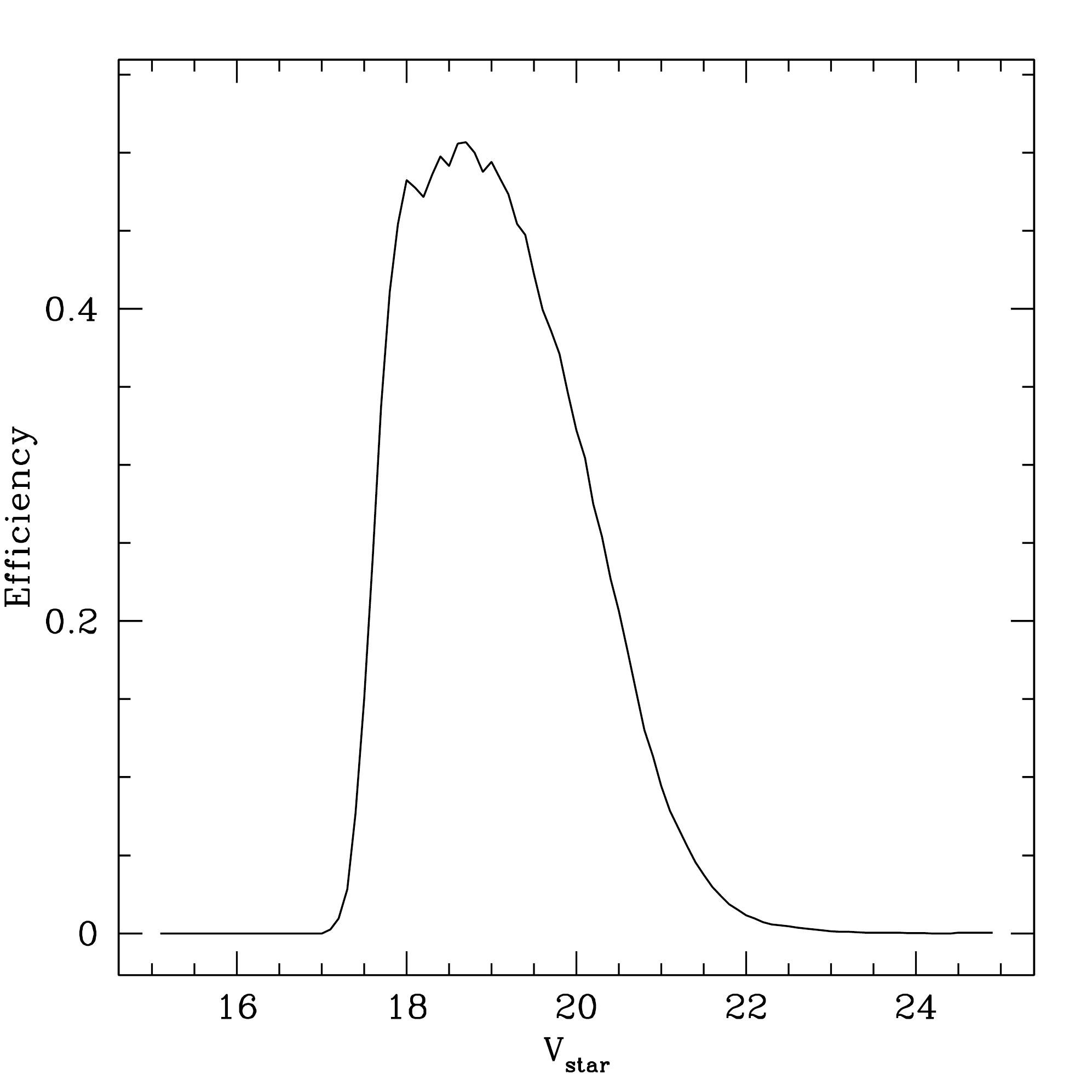}
\caption{The MACHO detection efficiency as a function of stellar V-magnitude.
If a microlensing event occurs in a \textit{star} of given magnitude
$V_{star}$, this is the given efficiency for detecting that event.}
\label{eff}
\end{figure}

Each CMD now contains a fraction $1.0-\fbkg$ source stars from the LMC
disk+bar and a fraction $\fbkg$ of source stars from the background
population. The CMD now represents the distribution of source stars down to
$V\sim24$.  However, not all possible microlensing events are detected in the
MACHO images.  To create a CMD representing a population of source stars which
produce detectable microlensing events we must convolve the HST CMD with the
MACHO detection efficiency. The MACHO efficiency pipeline is extensively
described by \citet{macho_lmc5.7} and \citet{macho_eff} and the detection
efficiency as a function of stellar magnitude, $V_{\rm{star}}$, and Einstein
ring crossing time has been calculated.  We average this function over the
event durations of the thirteen events from \citet{macho_lmc5.7} and
present the MACHO detection efficiency as a function of $V_{\rm{star}}$ in
Figure~\ref{eff}.  We convolve this function with our HST CMDs to produce the
final model source star CMDs.  

\begin{figure}.
\epsscale{0.5}
\plotone{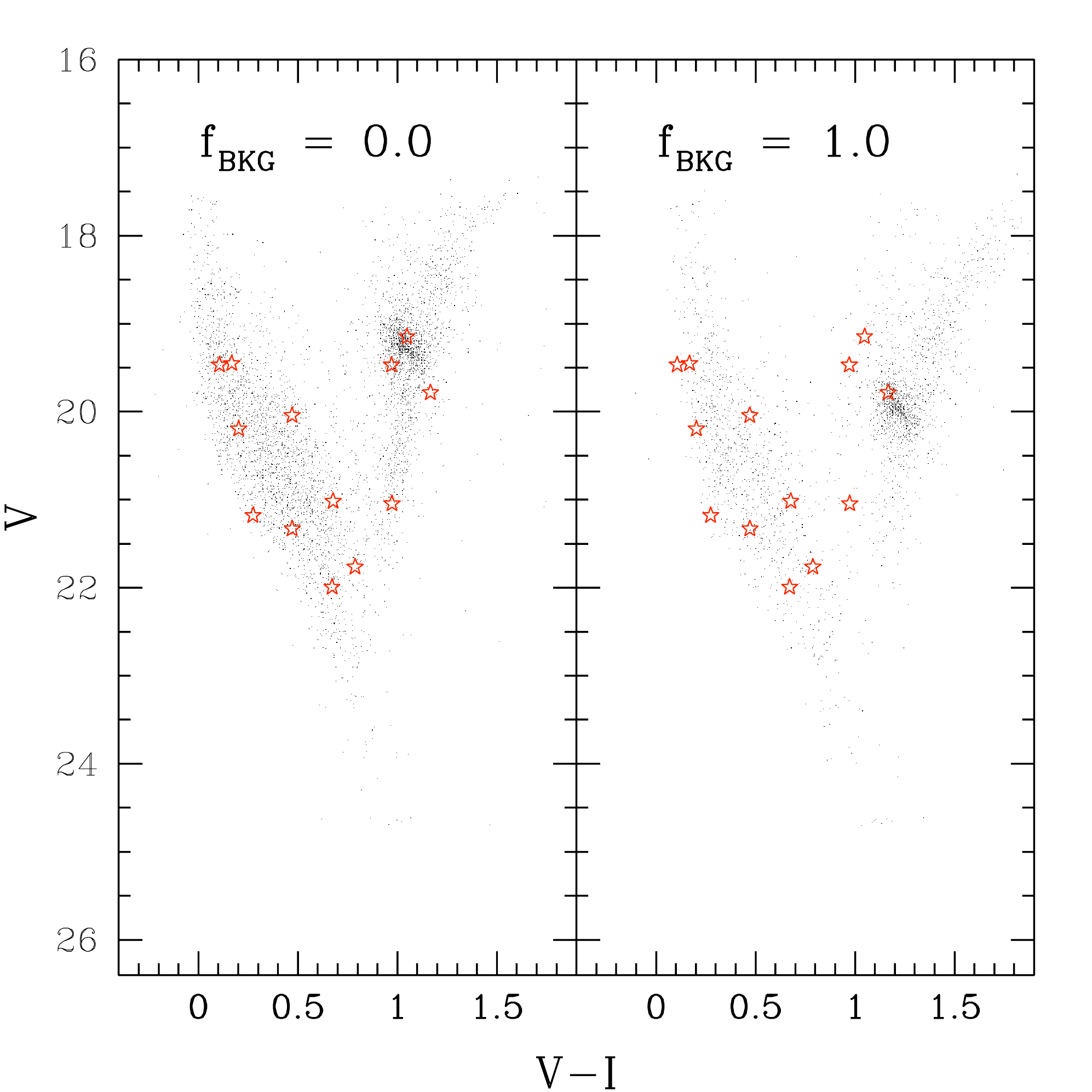}
\caption{
We show the observed microlensing source stars as large red
stars, overplotted on two model source star populations (small
black dots).  The left
hand panel represents a source star population drawn entirely
from the LMC disk+bar ($\fbkg=0.0$).  The right hand panel represents
a model in which all of the stars belong to a background population
($\fbkg=1.0$) with $\Delta \mu = 0.3$ and $\Delta E(V-I) = 0.18$.}
\label{model}
\end{figure}

In Figure~\ref{model}, we illustrate two model source star populations, both
with $\Delta \mu = 0.30$, one with $\fbkg = 0.0$ (all source stars in the LMC
disk+bar) and one with $\fbkg = 1.0$ and $\Delta \mu = 0.30$ (all source stars
in a background population displaced by 7.5 kpc). We overplot the observed
microlensing source stars of Table 2 as large red stars.

This procedure is based on several assumptions.  First, we assume that our
thirteen 
HST fields collectively well represent the stellar population of the LMC disk.
This assumption has two parts, the first being that an observation at a random
line of sight in the LMC bar is dominated by stars in the LMC disk and the
second that the stellar population across the LMC is fairly constant.  The
first part holds so long as the surface density of the background population
is much smaller than that of the LMC itself.  If this were not the case, this
population would have been directly detected. The second part has been
confirmed by many LMC population studies including \citet{macho_eff},
\citet{ols99}, and \citet{geh98}, as well as our own comparison of individual
CMDs and luminosity functions.  In addition, as mentioned above, we find
that all of our fields have reddenings within about $\Delta E(B-V) \sim 0.03$
of their mean value.

\section{Statistical Method and Results}
\label{stats}

We now determine which model source star population is most
consistent with the observed CMD of microlensed source stars by
performing a two-dimensional Kolmogorov-Smirnov test.

In the familiar one dimensional case, a KS test of two samples with number of
points $N_{1}$ and $N_{2}$ returns a distance statistic $D$, defined to be the
maximum distance between the cumulative probability functions at any ordinate.
Associated with $D$ is a corresponding probability $P(D)$ that \textit{if} two
random samples of size $N_{1}$ and $N_{2}$ are \textit{drawn from the same
distribution} a worse value of $D$ will result.  This is equivalent to saying
that we can exclude the hypothesis that the two samples are drawn from the
same distribution at a confidence level of $1.0-P(D)$. If $N_{2} \gg N_{1}$, then
this is also equivalent to excluding at a $1.0-P(D)$ confidence level the
hypothesis that sample 1 is drawn from sample 2.

The concept of a cumulative distribution is not defined in more than one
dimension.  However, it has been shown that a good substitute in two
dimensions is the integrated probability in each of four right-angled
quadrants surrounding a given point (Fasano \& Franceschini 1987; Peacock
1983).  That is, consider a point $(x_1,y_1)$ in distribution $N_1$ and draw
two perpendicular lines through this point (one line perpendicular to the $x$
axis and the other perpendicular to the $y$ axis) which divide the space into
four quadrants.  The integrated probability of each quadrant for each
distribution is the fraction of data from the distribution which lies in that
quadrant.  The two dimensional KS statistic $D$ is now taken to be the maximum
difference (ranging both over all data points and all quadrants) between the
integrated probability of distribution $N_1$ and $N_2$.  See Press \etal(1992)
for further details.  The distance statistic $D$ and the corresponding $P(D)$
are subject to the same interpretation as in the one dimensional case.  

\begin{figure}
\epsscale{0.5}
\plotone{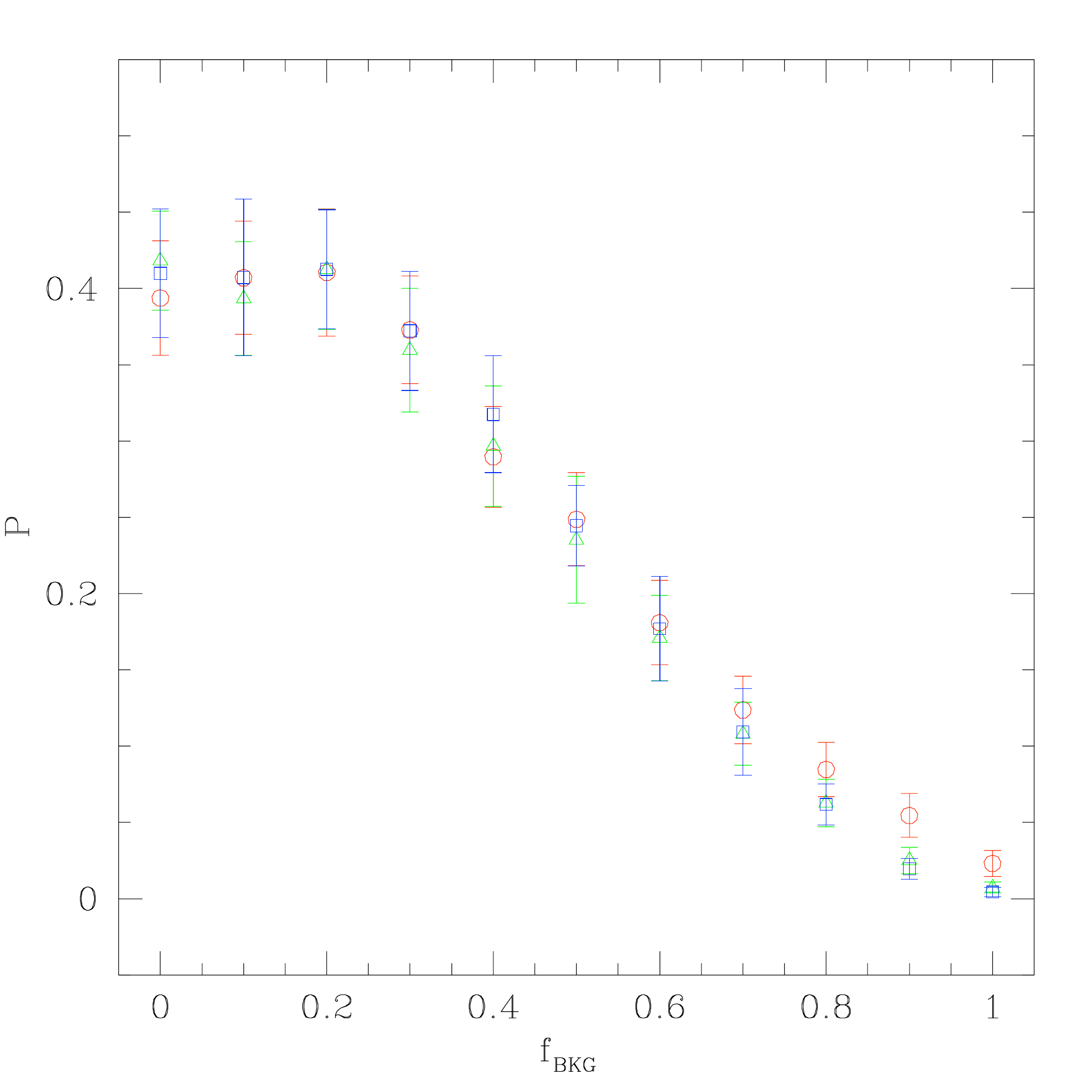}
\caption{Kolmogorov-Smirnov test results.
The 2-D KS test probability $P$ that our observed distribution of
source stars was drawn from a source star population in which a fraction
$f_{BKG}$ of the source stars are located behind the LMC.  The distance moduli
of the background stars with $\Delta \mu = 0.0, 0.30, 0.45$ are shown as red
circles, green triangles and blue squares respectively.  The errors bars
indicate the scatter about the mean value for 20 simulations of each model.
}
\label{results}
\end{figure}

We show the resulting $P(D)$ plotted versus $\fbkg$ in Figure~\ref{results}.
The results are shown separately for $\Delta \mu = 0.0, 0.3$ and $0.45$ in red
circles, green triangles, and blue squares, respectively. The error bars
indicate the scatter about the mean value for 20 simulations of each model.
Because the creation of the efficiency convolved CMD is a weighted random draw
from the HST CMD, the model population created in each simulation differs
slightly.  This in turn leads to small differences in the KS statistics.  We
find that the 2-D KS-test probability is highest at a fraction of source stars
behind the LMC $\fbkg ~ \sim 0.0 - 0.2$ with very little dependence on the
value of the displacement, $\Delta \mu$.  The significant preference for low
$\fbkg$ arises mostly from reddening/color.

In a strict statistical sense, the
KS test can only be used to reject or confirm the hypothesis that two datasets
have a common parent distribution.  As a general guideline, a probability of
less than 10\% suggests that the two distributions are significantly different
(Press \etal1992).  It is not strictly allowable to compare relative
probability, e.g., a KS test probability of 80\% does not mean that it is
twice as likely that two distributions are identical than if they had returned
a KS test probability of 40\%.  However, we still find it valuable to show 
the shape of our KS test probability function to demonstrate its insensitivity
to $\Delta \mu$.

\section{Discussion}
\label{ssdiscuss}
\label{2.6}

We now compare the observational results for $\fbkg$ (Figure 10) with the
expected value for our four models of microlensing: a) background lensing
($\fbkg\sim1.0$), b) LMC spheroid self-lensing ($\fbkg\simlt 0.65$), c) LMC
disk+bar self-lensing ($\fbkg\sim0.0$), and d) MW lensing ($\fbkg\sim0.0$).

We rule out a model in which the source stars all belong to some background
population at a confidence level of $99$\% (e.g., the KS test probability
for $\fbkg = 1.0$ is $<0.01$). Of course, one event of this sample, LMC-5, 
is known to be caused by a lens in the Milky Way disk, but $\fbkg = 0.9$ and
$\fbkg = 0.8$ are excluded by $>95\,$\% and $>90\,$\%, respectively. So, 
background source models would seem to be excluded. But, this conclusion
might be weakened if we had allowed for patchy extinction within the LMC.

We can rule out spheroid self-lensing models where the spheroid has the
same luminosity function as the disk+bar
($f_{BKG}\sim 0.65$) at the
statistically marginal confidence level of $80-90$\%. But old spheroid self-lensing
models without A and F stars in the spheroid have $f_{BKG} < 0.65$ and are
not significantly disfavored by this analysis. The relatively low number density
of LMC halo RR Lyrae \citep{lmc_stellar_halo} suggests that the LMC spheroid 
microlensing optical depth should be quite low, but there are several reasons not
to dismiss the LMC spheroid self-lensing models. First, as \citet{wei00} emphasized,
the LMC is not an isolated galaxy, and the interactions of the LMC with the Milky Way
and the Small Magellanic Could (SMC) significantly complicate attempts to properly model
its structure. Thus, the lack of an LMC model that provides a sufficient microlensing
optical depth while fitting the other observational constraints may just reflect our
failure to model the MW + LMC + SMC system. 

The allowed region of the $\fbkg$ plot ($P>10$\%) is consistent with the
expected location of the source stars in both the MW-lensing and LMC disk+bar
self-lensing geometries.  However, detailed modelling of LMC disk+bar
self-lensing suggests that it contributes only 25\% of the observed optical
depth \citep{gyu00,mancini_lmcdisk}.  So, the test presented here seems
most consistent with having the majority of events located in the MW halo, with some
contribution from the LMC disk+bar and MW disk.

There are, however, additional reasons that favor the LMC spheroid self-lensing model.
The strongest point in favor of LMC spheroid self-lensing is the apparent
spacial distribution of the LMC microlensing optical depth. While the 
spacial distribution of the MACHO events does not match expectations from
LMC disk+bar self-lensing \citep{mancini_lmcdisk}, LMC spheroid self-lensing
should produce a much more uniform distribution of events across the central
regions of the LMC where the MACHO survey had high sensitivity 
(within $\simlt 3^\circ$ of the LMC-bar center).  However, the
EROS survey has very low sensitivity to microlensing in this central part of the
LMC, as most of their sensitivity comes from fields that are $\simgt 3^\circ$ from
the center of the LMC-bar. Since LMC spheroid self-lensing should provide a much
lower optical depth in these outer fields, it would seem that LMC spheroid
self-lensing might provide a reasonable explanation for both the MACHO and
EROS results.

Some of the handful of Magellanic Cloud microlensing events with distance 
estimates also provide some clues to the locations of the other lenses.
LMC-5 and LMC-20 (from the B sample of \citet{macho_lmc5.7}) have both
been confirmed to be MW disk lenses \citep{lmc5-detect,kal_lmc20}, but 
they have been detected by methods that can only detect lenses that are close
to us, so they don't really inform us about the location of other events. The
situation is similar for the binary source event, LMC-14, which has been shown 
to be caused by a lens in the LMC disk+bar or spheroid \citep{96lmc2}. The
binary source effects that lead to this conclusion are only detectable for 
lenses that reside in the LMC, some self-lensing events are expected for any
reasonable model of the LMC. On the other hand, only a minority of self-lensing
events should show this effect, so one might argue that such an event should be
accompanied by a number of other self-lensing events that don't show this effect.
But this argument doesn't have much statistical significance since it is based on
only a single event.

It is more informative to consider events that can reveal distance information
no matter where the lens resides. There are two types of such events: binary lens
events with resolved caustic crossings, and events with space-based microlensing
parallax measurements. Event LMC-9 is a binary lens event with a caustic crossing
that is marginally resolved by only two measurements \citep{macho-lmc9}. The
long caustic crossing time favors a LMC disk+bar lens, but due to the poor sampling
of the caustic crossing alternative interpretations are possible. Event MACHO-98-SMC-1
was announced by the MACHO Alert system \citep{macho-alert,98smc1-disc}
prior to the second caustic crossing, which enabled the EROS, MPS, OGLE and
PLANET collaborations to join MACHO in monitoring the event. This allowed
the second caustic crossing to be well resolved, and the observed caustic crossing
timescale indicated that the lens resided in the SMC \citep{joint-98smc1}. However,
the SMC is known to be much more extended along the line-of-sight than the LMC,
so its self-lensing optical depth is expected to be much larger. So, the fact that
this lens is likely to be in the SMC does not imply that most LMC sources have
lenses that are also in the LMC.

The situation with the only event (OGLE-2005-SMC-1)
with a space-based microlensing parallax
measurement \citep{ogle05smc1} is somewhat more interesting because the
analysis favors a MW halo lens over a lens in the SMC by a 20:1 likelihood ratio.
However, if the lens is in the MW halo, then it is almost certainly more massive
than the lenses for the LMC microlensing events found by MACHO. So, it doesn't
really support the MW halo model for the LMC lenses, either. However, it does
seem plausible that this event could be explained by a SMC-spheroid lensing
population that is larger than predicted by standard SMC models.

\section{Summary}
\label{sssum}

We have presented HST follow-up observations of 13 LMC microlensing
events presented in \citet{macho_lmc5.7} including the 12 genuine microlensing
events passing the cuts for for selection criteria A. These HST data support
our conclusions, based on light curve modeling, that these are genuine microlensing
events, and they also strengthen the conclusion that the MACHO LMC microlensing
sample does not have significant contamination due to non-microlensing variability.

We have also compared the colors and magnitudes of our source stars
to the predictions of various models to determine which models of LMC
microlensing might be consistent with the observed events. 
At present, the strength of this analysis is severely limited by the number of
microlensing events.  We are currently able to exclude the most
extreme model ($\fbkg\sim1.0$), but with more events, it should be possible
to more accurately determine $\fbkg$ and judge 
intermediate models such as LMC spheroid or
shroud self-lensing. Ongoing microlensing
search projects (OGLE-III, MOA-II, SuperMACHO) 
may supply a sufficient sample of events in the next few
years.  The technique outlined in this paper should prove a powerful method
for locating the sources, and, subsequently, the lenses with these future datasets.

At present, there exist no theoretical models that adequately explain the microlensing
optical depth seen towards the LMC.  LMC spheroid or shroud models seem 
most likely to fit the spacial distribution of microlensing events seen in
the combined MACHO plus EROS data sets.

\acknowledgments
The work of CAN, AJD, KHC, PP. MG, SCK, and SLM work was performed under
the auspices of the U.S. Department of Energy by Lawrence Livermore
National Laboratory in part under Contract W-7405-Eng-48 and in part under
Contract DE-AC52-07NA27344.
D.P.B.\ was supported by grant
AST-0708890 from the NSF.
D.M.\ is supported by the Basal CATA, by FONDAP Center for Astrophysics 
15010003,
and by the Milky Way Millennium Center.

{\it Facilities:} \facility{MtS:1.3m (MACHO Camera)}, \facility{HST (WFC2)}

\appendix

\clearpage

\clearpage

\begin{deluxetable}{lllcccl}
\tablewidth{0pt}
\tablecaption{WFPC2 Observations of Microlensing Source Stars}
\tablehead{
\colhead {Event} & \colhead{RA} & \colhead {DEC} & \colhead {V} & \colhead{R}
&\colhead{I} &\colhead{Obs Date}}
\startdata
LMC-1 & 05:14:44.53 & $-$68:47:56.31 &   $4~\times~400$ & $4~\times400$ & $40~\times~400$ & 1997/12/16\\
LMC-4 & 05:17:14.59 & $-$70:46:55.28 &   $3~\times~500$ & $3~\times500$ & $33~\times~500$ & 1997/12/12\\
LMC-5 & 05:16:41.22 & $-$70:29:21.22 &   $4~\times~400$ & $2~\times400$ & $2~\times~400$ & 1999/05/13\\
LMC-6 & 05:26:13.77 & $-$70:21:16.37 &   $4~\times~400$ & $2~\times400$ & $2~\times~400$ & 1999/08/26\\
LMC-7 & 05:04:03.61 & $-$69:33:21.00 &   $4~\times~400$ & $2~\times400$ & $2~\times~400$ & 1999/04/12\\
LMC-8 & 05:25:09.65 & $-$69:47:53.82 &   $4~\times~400$ & $2~\times400$ & $2~\times~400$ & 1999/03/12\\
LMC-9 & 05:20:20.52 & $-$69:15:13.84 &   $4~\times~400$ & $2~\times400$ & $2~\times~400$ & 1999/04/13\\
LMC-13 & 05:24:02.83 & $-$68:49:14.77 &  $3~\times~500$ & $3~\times500$ & $3~\times~500$ & 2000/07/28\\
LMC-14 & 05:34:44.54 & $-$70:25:10.09 &  $4~\times~500$ & $2~\times400$ & $4`\times~500$ & 1997/05/13\\
LMC-15 & 05:05:45.86 & $-$69:43:54.02 &  $3~\times~500$ & $3~\times500$ & $3~\times~500$ & 2000/07/17\\
LMC-18 & 05:45:20.87 & $-$71:09:15.20 &  $3~\times~500$ & $3~\times500$ & $3~\times~500$ & 2000/07/21\\
LMC-20 & 04:54:18.81 & $-$70:02:21.39 &  $3~\times~500$ & $3~\times500$ & $3~\times~500$ & 2000/07/29\\
LMC-21 & 04:57:13.82 & $-$69:27:50.57 &  $3~\times~500$ & $3~\times500$ & $3~\times~500$ & 2000/07/26\\
LMC-23 & 05:06:16.85 & $-$70:58:49.98 &  $3~\times~500$ & $3~\times500$ & $3~\times~500$ & 2000/07/18\\
LMC-25 & 05:02:15.86 & $-$68:00:55.10 &  $3~\times~500$ & $3~\times500$ & $3~\times~500$ & 2000/07/14\\
\enddata
\tablecomments{The columns V and I indicate the number of single exposures
times the number of seconds per exposure.}
\end{deluxetable}

\begin{deluxetable}{lrrrrrr}
\tablewidth{0pt}
\tablecaption{WFPC2 and MACHO Blend Fit Photometry of Microlensing Source Stars}
\tablehead{
\colhead {Event} & \colhead{$V_{\rm HST}$} & \colhead {$R_{\rm HST}$} & \colhead{$I_{\rm HST}$} &
\colhead{$V_{\rm blend}$} & \colhead{$R_{\rm blend}$} &\colhead{$\chi^2$}
}
\startdata
LMC-1 & $19.78\pm0.03$ & $19.20\pm0.03$ & $18.61\pm0.03$ & $19.74\pm0.02$ & $19.19\pm0.02$ &  1.510 \\ 
LMC-4 & $21.33\pm0.03$ & $21.09\pm0.03$ & $20.83\pm0.03$ & $21.31\pm0.26$ & $21.08\pm0.26$ &  1.560 \\ 
LMC-5 & $21.02\pm0.07$ & $20.72\pm0.07$ & $20.34\pm0.08$ & $21.27\pm0.02$ & $21.01\pm0.03$ &  0.940 \\ 
LMC-6 & $20.04\pm0.03$ & $19.86\pm0.03$ & $19.57\pm0.03$ & $19.95\pm0.22$ & $19.82\pm0.23$ &  0.850 \\ 
LMC-7 & $21.99\pm0.03$ & $21.76\pm0.04$ & $21.32\pm0.03$ & $20.93\pm0.02$ & $20.83\pm0.03$ &  1.410 \\ 
LMC-8 & $20.20\pm0.03$ & $20.18\pm0.03$ & $19.99\pm0.03$ & $20.35\pm0.10$ & $20.09\pm0.10$ &  3.280 \\ 
LMC-9$^{*}$ & $21.14\pm0.03$ & $20.65\pm0.03$ & $20.15\pm0.03$ & $21.43\pm0.02$ & $20.88\pm0.02$ &  1.490 \\ 
LMC-13 & $21.76\pm0.03$ & $21.38\pm0.03$ & $20.92\pm0.03$ & $22.57\pm0.43$ & $22.11\pm0.43$ &  1.210 \\ 
LMC-14 & $19.47\pm0.03$ & $19.46\pm0.03$ & $19.36\pm0.03$ & $19.43\pm0.08$ & $19.46\pm0.08$ &  0.810 \\ 
LMC-15 & $21.18\pm0.03$ & $21.07\pm0.03$ & $20.90\pm0.03$ & $21.48\pm0.73$ & $21.37\pm0.79$ &  0.940 \\ 
LMC-18 & $19.47\pm0.03$ & $18.99\pm0.03$ & $18.50\pm0.03$ & $19.58\pm0.17$ & $19.31\pm0.17$ &  1.290 \\ 
LMC-21 & $19.45\pm0.03$ & $19.39\pm0.03$ & $19.28\pm0.03$ & $20.11\pm0.17$ & $20.08\pm0.17$ &  2.580 \\ 
LMC-23 & $21.05\pm0.03$ & $20.64\pm0.03$ & $20.07\pm0.03$ & $20.31\pm0.04$ & $20.18\pm0.05$ &  1.480 \\ 
LMC-25 & $19.15\pm0.03$ & $18.61\pm0.03$ & $18.10\pm0.03$ & $19.10\pm0.07$ & $18.58\pm0.05$ &  0.960 \\ 
\enddata
\tablecomments{The first three columns give the $V$, $R$, and $I$ WFPC2
magnitudes.  The errors include the photon-counting errors reported by DAOPHOT
II as well as an additional contribution of 0.03 mag of uncertainty due to
aperture corrections. The LMC-5 HST photometry includes the PSF fitting photometry
errors returned by ALLSTAR.  The columns $V_{\rm blend}$ and $R_{\rm blend}$ give
the calibrated lensed flux from a microlensing blend fit to the
ground-based
light curve. The $\chi^2$ column reports the reduced $\chi^2$ of the
microlensing blend fit.\\
$^{*}$ LMC-9 is believed to be an event in which the lens is a binary object.
The data presented here are for a binary blend-fit to the lightcurve.}
\end{deluxetable}

\end{document}